\documentstyle[epsfig]{aipproc}

\begin{document}
\title{Hard X-ray Emission from\\ Cassiopeia A SNR}

\author{Lih-Sin The$^*$, Mark D. Leising$^*$, Dieter H. Hartmann$^*$,\\ 
James D. Kurfess$^{\dagger}$, Philip Blanco$^{\S}$,
and Dipen Bhattacharya$^{\ddagger}$}
\address{$^*$ Department of Physics and Astronomy, Clemson University,
Clemson, SC 29634-1911 \\
$^{\dagger}$E.O. Hulburt Center for Space Research, Naval Research 
Laboratory,\\ Code 7650, Washington, DC 20375-5352 \\
$^{\S}$Center for Astrophysics and Space Sciences, University of
California, \\
San Diego, CA 92093\\
$^{\ddagger}$Institute of Geophysics and Planetary Sciences,
University of California,\\
Riverside, CA 92521}

\maketitle

\begin{abstract}
  We report the results of extracting the hard X-ray continuum spectrum
  of Cas A SNR from RXTE/PCA Target of Opportunity observations (TOO) and
  CGRO/OSSE observations. 
  The data can rule out the single thermal bremsstrahlung model for
  Cas A continuum between 2 and 150 keV.
  The single power law model gives a mediocre fit ($\sim$5\%) to the 
  data with a power-law index, $\Gamma$ = 2.94$\pm$0.02.
  A model with two component (bremsstrahlung + bremsstrahlung or
  bremsstrahlung + power law) gives a good fit.
  The power law index is quite constrained suggesting
  that this continuum might not be the X-ray 
  thermal bremmstrahlung from accelerated MeV electrons
  at shock fronts\cite{asvar90}  
  which would have $\Gamma\simeq$2.26.
  With several SNRs detected by ASCA 
  showing a hard power-law nonthermal X-ray continuum,
  we expect a similar situation for Cas A SNR which has 
  $\Gamma$=2.98$\pm$0.09.
  We discuss the implication of the hardest nonthermal X-rays detected
  from Cas A to the synchrotron radiation model.
\end{abstract}

\section*{Introduction}
      X-ray observations of supernova remnants have been stimulated by
  recent reports of nonthermal power-law X-ray detections
  suggesting supernovae as sites of charged particle accelerations and
  sources of cosmic rays.
  The first strong evidence for charged particle acceleration near
  supernova shock fronts in the X-ray energy band is demonstrated by the 
  morphological and spectral correlation between  X-ray and 
  radio emission from the bright NE and SW rims of SN 1006 \cite{koyam95}. 
  The brightest radio emission regions show almost featureless
  power law X-ray spectra when compared with SN 1006 central region which
  is dominated by emission lines of highly ionized elements in a 
  non-equilibrium ionization thermal plasma.
  The nonthermal X-ray component 
  has been modeled as a synchrotron emission from
  electrons accelerated to $\sim$100 TeV within shock fronts 
  \cite{ammov94,reyno96,masti96}.
  In this model, the radio spectrum is produced by synchrotron
  radiation of electrons accelerated to $\sim$GeV energies
  with the radio power-law index being less steep than 
  the X-ray power-law index.
    Several similar evidences also have been demonstrated by
  ASCA measurement of RX J1713.7-3946 \cite{koyam97} and
  IC 443 \cite{keoha97}, and 
  RXTE and OSSE measurements of Cas A \cite{roths97,allen97,the96}.

  OSSE with a total accumulation time of 15$\times$10$^5$ detector-seconds,
  detected a hard continuum between 40-150 keV from Cas A SNR at
  a 4$\sigma$ confidence with a flux of 
  $\sim$9$\times$10$^{-7}$ $\gamma$ cm$^{-2}$ s$^{-1}$ keV$^{-1}$ 
  at 100 keV \cite{the96}.
  The detection is the 
  hardest X-ray detection from a SNR without plerionic source.
  However, the shape of the continuum has not been strongly constrained.
  The continuum can either be a bremsstrahlung with kT$\simeq$35 keV or 
  a power law with $\Gamma\simeq$3.06.
   In this paper, we use the 2-30 keV RXTE/PCA TOO and 
  the 40-150 keV CGRO/OSSE Cas A data to better determine the shape
  of Cas A hard X-ray continuum.

%---------------------------------------------------------------------
\begin{figure}[t]
\centerline{\epsfig{file=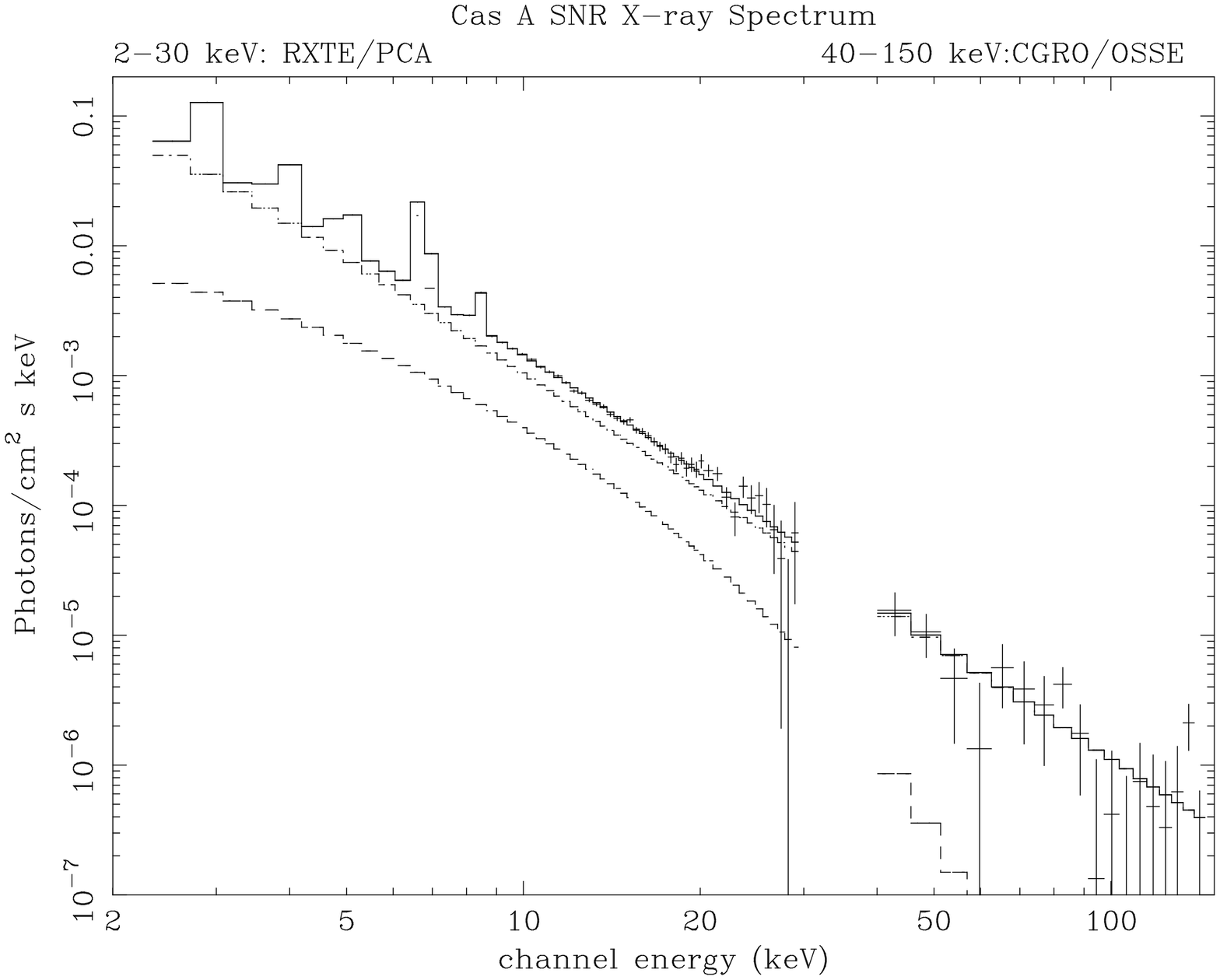,height=7.3cm,bbllx=545pt,bblly=79pt,bburx=70pt,bbury=665pt,angle=-90}}
\caption{The thermal bremsstrahlung (kT=7.93 keV; dashed line) + 
         power-law ($\Gamma$=2.98; dashed-dotted line) + 6 K X-ray line model 
         produces a good fit (solid line) to the PCA \& OSSE data (crosses)
         with a $\chi^2$/dof=1.077, dof=62.}
\label{bremspwrl}
\end{figure}
%---------------------------------------------------------------------

%---------------------------------------------------------------------
\begin{figure}[t] % fig 1
\centerline{\epsfig{file=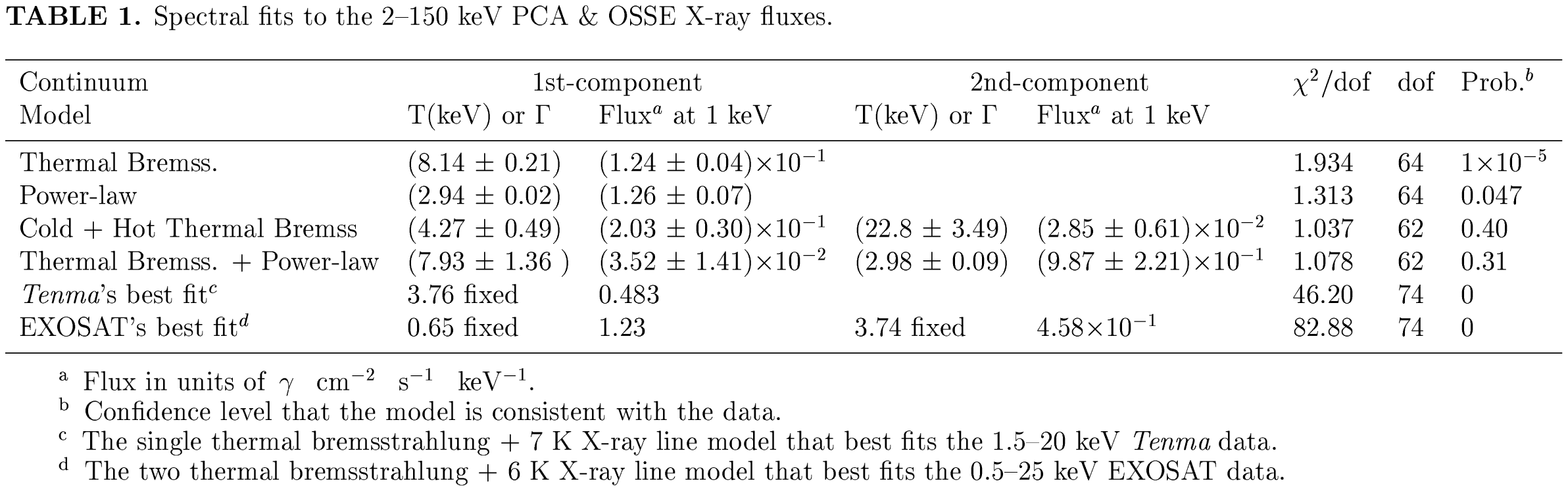,width=14cm,bbllx=545pt,bblly=34pt,bburx=324pt,bbury=723pt,angle=90}}
\label{fig1}
\end{figure}
%______________________________________________________________________________

\section*{Results}
 
  The RXTE/PCA data used in this analysis is the TOO by RXTE on 
  Aug. 2, 1996 with a total observing time of $\sim$4000 sec.
  We fit the 2-30 keV PCA data simultaneously with the 40-150 keV OSSE
  data. 
  The results of fitting several models with one or two continuum plus 
  about six emission lines from Si, S, Ar, Ca, and Fe are shown in Table 1.
  The line widths are fixed to 100 eV and
  the hydrogen column density toward Cas A is fixed at 
  1$\times$10$^{22}$ H cm$^{-2}$\cite{tsune86,jansen88}.
  We find that a single thermal bremsstrahlung model for the data 
  can be ruled out.
  Nevertheless, a single power law model gives a mediocre fit.
  We also find the two models that best fit the EXOSAT\cite{jansen88} 
  and {\it Tenma}\cite{tsune86} data
  cannot give acceptable fits to the PCA and OSSE data.
  A two component model, either the bremsstrahlung + bremsstrahlung or
  the bremsstrahlung + power law, each gives an equally good fit.
  In Figure 1 for the current interest in a nonthermal power law model, 
  we show the result of the thermal bremsstrahlung + power-law model 
  fit to the data. 
  The power law continuum is detected with a higher confidence level and 
  the power-law index is more tightly constrained than
  using the OSSE data alone\cite{the96}. 
  The PCA and OSSE data is consistent only at a 3.8\% level with the 
  X-ray bremsstrahlung from accelerated MeV electrons at shock fronts
  as suggested by Asvarov et al.\cite{asvar90} which would have 
  $\Gamma=\alpha$+1.5=2.26 where $\alpha$ is the radio spectral index.

\section*{Discussions}
   RXTE PCA + HEXTE in its AO-1 observing period detected a continuum 
   between 10 and 60 keV\cite{roths97,allen97}.
  The continuum is fitted well with a model of 
  two Raymond-Smith thermal bremsstrahlung
  of kT = 0.7 keV and kT = 2.8 keV 
  and a power law with $\Gamma\sim$2.4 which exponentially steepens at
  e-fold 50 keV.
  The shape of the model is in good agreement with 
  the result we find here.

  The bremsstrahlung + bremsstrahlung model is expected in 
  the reverse shock model.
  In this model,
  the hot temperature (T$\sim$23 keV) of the shocked circumstellar
  gas behind the blast-wave is consistent with the assumption that the
  blast-wave velocity is slightly larger than the observed median-filament
  velocity of 5500 km/s\cite{woltj72,mckee74} and the circumstellar
  density $n_H<$1.5 cm$^{-3}$.
  The shocked gas temperature behind the reverse shock of T$\sim$4 keV
  suggests the swept up mass is $\sim$6 M$_{\odot}$ and 
  $n_H\simeq$1.4 cm$^{-3}$\cite{mckee74,borko96}.

%---------------------------------------------------------------------
\begin{figure}[t]
\centerline{\epsfig{file=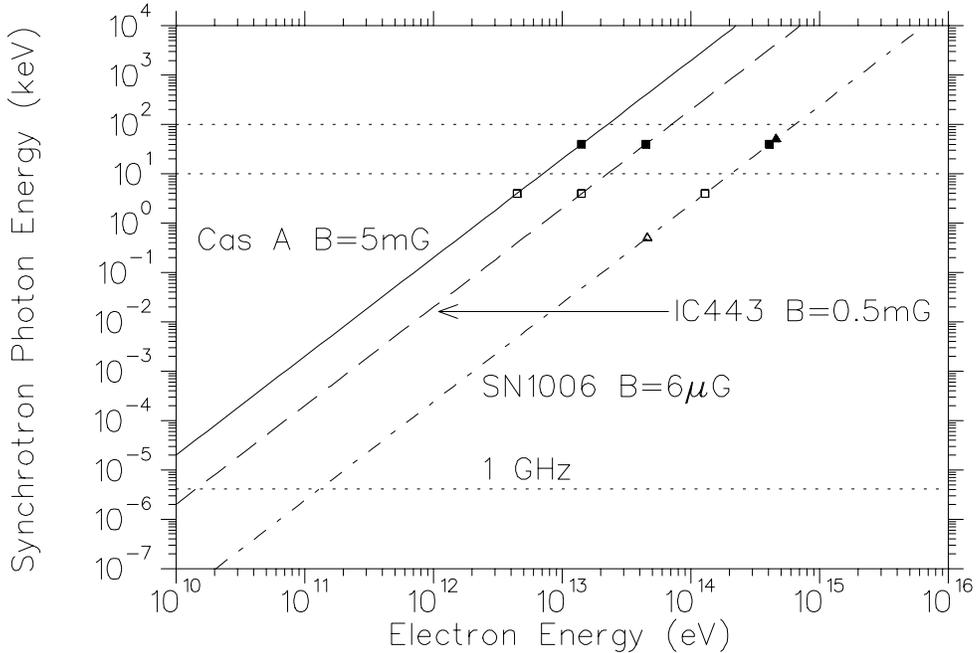,width=13cm,bbllx=90pt,bblly=400pt,bburx=500pt,bbury=675pt}}
\caption{The synchrotron photon energy relation with the 
         electron energy \protect\cite{koyam95} 
         for Cas A (solid line), IC 443 (dashed line), and 
         SN 1006 (dashed-dotted line).
         The solid squares and the open squares mark 
         the maximum electron energy constrained by
         the synchrotron energy loss (Eq.(1) of \protect\cite{reyno96})
         for f*R$_J$=1 and f*R$_J$=10, respectively.
         The solid triangles and the open triangles mark 
         the maximum electron energy bounded by the age of the SNRs
         (Eq.(2) of \protect\cite{reyno96})
         for f*R$_J$=1 and f*R$_J$=10,
         respectively.
         This maximum energy is shown only for SN 1006 due to
         its low magnetic field.
         Synchrotron energy loss sets the maximum electron energy in
         Cas A and IC 443.
         The shock speed is assumed to be 5000 km/s in each case.
         f is the ratio of the electron mean free path to 
         the electron gyroradius.
         R$_J$ is the factor that contains the orientation of the shock 
         relative to the magnetic field. 
         The dotted lines shows some interesting 
         synchrotron photon energies at 
         1 GHz, 10 keV, and 100 keV.}
\label{FigSynchrotron}
\end{figure}
%---------------------------------------------------------------------

   A summary of the detected nonthermal continuum from four SNRs 
   is shown in Table 2.
   Cas A is of interest because from the 4 SNRs with measured nonthermal
   power-laws, it is the youngest, has the strongest magnetic field, 
   its detected nonthermal energy is the highest, 
   and it has the steepest X-ray and radio power-law indexes, 
   of four SNRs with measured nonthermal power law.
   In Figure 2, we show for Cas A, IC 443, and SN 1006
   how close their detected hard X-ray energies to 
   the maximum synchrotron energies.
   The synchrotron energy is related to
   the maximum electron energy which is limited 
   by the synchrotron energy loss and the SNR's age.
   Cas A, as the brightest radio SNR source, has the most intense magnetic
   field and therefore it causes the large synchrotron energy loss 
   which limits Cas A's maximum electron energy.
   Figure 2 shows that the hard X-ray emission detected from Cas A by OSSE
   is near the cutoff synchrotron energy or  indirectly to
   the maximum electron energy. 
   Further Cas A measurements by OSSE may detect the cutoff energy and
   hence would provide the evidence that the nonthermal power law emission
   is the synchrotron radiation from accelerated electrons at the shock
   fronts near the maximum energy.
   Another evidence of the synchrotron radiation being the process for the
   nonthermal emission is the  steepening of spectral index from radio
   to X-ray energy. It has been suggested that the turnover or
   break $\sim$0.25 keV is observed in SN 1006 spectra\cite{masti96}. 
   We expect the turnover in Cas A spectra is $\sim$1-10 keV due to
   Cas A's strong magnetic field.  
   However, in order to observe this turnover in Cas A spectra,
   an X-ray imaging detector is needed
   because Cas A's thermal bremsstrahlung contributes substantially 
   near this energy.

%-------------------------------------------------------------------
\setcounter{table}{1}
\begin{table}[t]
\caption{Comparison of the measured nonthermal continuum}
\label{table2}
\begin{tabular}{llcdld}
  SNR  & Age  & Best Estimate & $\alpha$ & 
                                  \multicolumn{2}{c}{Measured Power Law} \\ 
       & (yrs) &    of B      &          &  Energy (keV)   &   $\Gamma$  \\
\noalign{\smallskip}
 \hline
\noalign{\smallskip}
 SN1006          & 971         & 3-10 $\mu$G  &  0.56 & 0.5 - 20 & 2.95 \\
 RX J1713.7-3946 & $\geq$1000  &   ?          &   ?   & 0.5 - 10 & 2.45  \\
 IC 443          & 1000        & 500 $\mu$G  & $<$0.24 & 0.5 - 12 & 2.35 \\
 Cas A           & 317         & 1-5 mG       &  0.76 &  15-150 & 2.98 \\
\noalign{\smallskip}
\hline
\end{tabular}
\end{table}

\end{document}